\documentclass[dvips,a4paper]{article}

\usepackage{icrc2011}

\newcommand{\degree}{\ensuremath{^\circ}}

\title{Very high energy follow-up observations of GRBs detected by Fermi and Swift}

\newcommand{\etal}{\MakeLowercase{\textit{et al. }}} 
\shorttitle{Aune \etal VHE Obs. of GRBs det. by Fermi and Swift}

\authors{T. Aune$^{1}$, for the VERITAS Collaboration$^{2}$ }
\afiliations{$^1$Department of Physics and SCIPP, 
University of California, Santa Cruz, CA 95064 U.S.A.\\ 
$^2$http://veritas.sao.arizona.edu/conferences/authors?icrc2011' }

\email{aune@physics.ucsc.edu}

\abstract{In many theoretical models of gamma-ray bursts (GRBs) and their afterglows, the emission of photons above 100 GeV is predicted. The Large Area Telescope (LAT) on-board the {\it Fermi} Gamma-ray Space Telescope has detected delayed, high-energy emission (up to 90 GeV in the burst rest-frame) from several GRBs and no evidence of a high-energy spectral cutoff during the early afterglow phase of the burst has been found. Presented here are the results of follow-up observations with VERITAS, a ground-based telescope array sensitive to gamma-rays above 100 GeV, of GRBs detected by the {\it Fermi} and {\it Swift} satellites. These observations have not yielded a conclusive detection and the upper limits on very high energy (VHE, E$>$100 GeV) gamma-ray flux obtained from these observations are among the most constraining to date.}

\begin{document}
\maketitle

\section{Introduction}

Observations of gamma-ray bursts (GRBs) and their afterglows are generally consistent with the relativistic fireball framework \cite{1999PhR...314..575P}. In this framework, the prompt gamma-ray emission is produced by internal shocks created by collisions of relativistic jets with differing Lorentz factors which originate from a central engine. The afterglow emission is produced, at least in part, by the external shocks created when the outflowing material interacts with the surrounding environment. Within this scenario high energy (HE, E$>$100 MeV) photons may be produced by both synchrotron and inverse Compton (IC) processes. 

These external shocks may also be capable of producing VHE gamma rays and characterizing this emission can directly constrain the medium density and the equipartition fraction of the magnetic field in the GRB environment \cite{2005ApJ...633.1018P}. The VHE emission from synchrotron self Compton processes in the forward shock may be delayed with respect to the low-energy gamma-ray emission by hundreds to thousands of seconds \cite{2008MNRAS.384.1483F}. Though this component is expected to be sub-dominant with respect to the synchrotron emission (and therefore difficult to detect with the Large Area Telescope (LAT) on-board {\it Fermi}), the very high energy and large delay of this component, make it a prime candidate for detection by ground-based imaging atmospheric Cherenkov telescope (IACT) systems \cite{2009ApJ...703...60X}.

Another mechanism for generating delayed VHE emission from GRBs is the IC scattering of photons from X-ray flares. The X-ray Telescope (XRT) on-board the {\it Swift} satellite has made it possible to take detailed X-ray observations of GRB afterglows on a regular basis and X-ray flare activity occurring hundreds to thousands of seconds after the initial burst has been detected in a large fraction of GRB afterglows \cite{2007ApJ...671.1903C}. A recent result from the {\it Fermi} collaboration \cite{FermiFlare} is the detection by the LAT of HE emission associated with X-ray flare activity in GRB\,100728A. The emission detected by the LAT extended to $>$1 GeV with a photon spectrum $\frac{dN}{dE} \propto E^{-1.4}$ observed hundreds of seconds after the initial burst, which makes the gamma-ray emission associated with X-ray flares a very interesting target for IACT observations.

Presented here are the results from GRB follow-up observations made over a $\sim 4$ year interval with the Very Energetic Radiation Imaging Telescope Array System (VERITAS) between 2007 and 2011. The sample includes GRBs detected by {\it Swift}, the {\it Fermi} Gamma-ray Burst Monitor (GBM) and one GRB detected by {\it Fermi}-LAT which was later found to be outside the VERITAS field of view (FOV). No significant source of VHE gamma rays was found to be associated with any of the observed GRBs. 

\section{VERITAS}

VERITAS is an array of four IACTs, each 12 m in diameter, located 1268 m a.s.l. at the Fred Lawrence
Whipple Observatory in southern Arizona, USA (31\degree 40' 30'' N, 110\degree 57' 07'' W). The
first telescope was completed in the spring of 2005 and the full, four-telescope array began routine
observations in the autumn of 2007. The first telescope was installed at a temporary location as a
prototype instrument and in the summer of 2009 it was moved to a new location in the array to make
the distance between telescopes more uniform and consequently improve the sensitivity of the system
\cite{2009arXiv0912.3841P}. The observations presented here were taken under moderate to good weather and with at least three of the four telescopes in the array operational. 

 \begin{figure*}[th]
  \centering
  \includegraphics[scale=0.9]{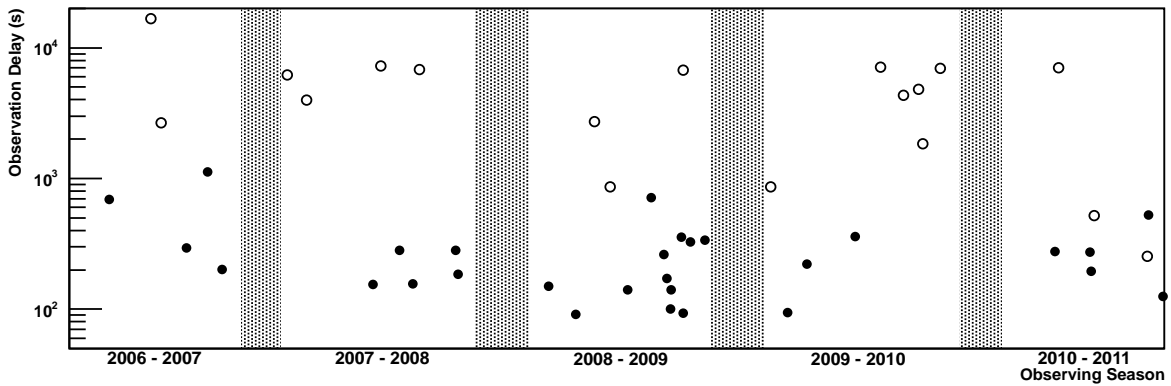}
  \caption{Delay from the start of the burst to the beginning of VERITAS observations for all GRBs
    with VERITAS data. The open symbols correspond to observations that were delayed due to
    constraints such as the burst occurring during daylight or below the horizon. Filled symbols are
    unconstrained observation delays and are primarily determined by the time it takes the
    telescopes to slew to the burst. The average unconstrained observation delay is four minutes. The shaded regions indicate the annual shutdown of the array due to the summer monsoons.}
  \label{grb_obs_delay}
 \end{figure*}

\section{GRB Observations}

GRB observations take priority over all other targets in the VERITAS observing plan. To allow for rapid follow-up observations of GRBs detected by satellites, the VERITAS control computers are set to receive notices from the GRB Coordinates Network (GCN)\footnote{http://gcn.gsfc.nasa.gov} and load the target positions automatically. The observers on duty are notified to stop the current observations and begin slewing the telescopes to the GRB position. The telescopes are capable of slewing at a rate of 1\degree/s in both azimuth and elevations. If the GRB position and time are not subject to observing constraints, the average delay from the beginning of the GRB to the beginning of VERITAS observations is 240 seconds. Figure \ref{grb_obs_delay} shows the observation delays for GRBs with VERITAS data. For well-localized GRBs, VERITAS observations will nominally continue for three hours after the burst, while, for bursts with positions obtained only from the {\it Fermi}-GBM which are poorly localized, VERITAS observations continue for one hour. 

As of this writing, VERITAS has taken follow-up observations of 49 GRBs. 47 of these were taken with at least three telescopes active in the array and 31 of the observed GRBs had positions determined by the {\it Swift}-XRT which has an angular localization well below the VERITAS point spread function (PSF). One observation (GRB\, 070311) was triggered by the INTEGRAL spacecraft, while the remaining fifteen were triggered by the {\it Fermi}-GBM. The GRB locations obtained from the GBM generally have $68\%$ containment radii of several degrees and it is not clear that these GRBs were in the FOV of VERITAS during these observations. For one burst, GRB\,110428A, VERITAS performed followed up observations of both the GBM flight and ground positions but the subsequent LAT detection found the burst to be located several degrees from the VERITAS pointing and outside the VERITAS FOV.

\section{Data Analysis}

The data obtained from GRB follow-up observations are analyzed using the standard VERITAS analysis suite \cite{2008ICRC....3.1385C}. For GRBs, the ''standard-source'' analysis is optimized for a weak source (3\% Crab flux) with a spectrum $\frac{dN}{dE} \propto E^{-\Gamma}$  where $\Gamma = 2.5$. For each burst, a ''soft-source'' analysis was performed as well which assumes a spectral index of $\Gamma = 3.5$. The spectra of GRBs above 100 GeV is unknown, but the standard analysis spectral index of 2.5 was chosen based on the average high-energy spectral index, $\beta$ observed by BATSE \cite{Kaneko:2006p3159}. It is expected that for any GRB with a redshift greater than $\sim 0.1$, the attenuation due to interaction of VHE photons with the extragalactic background light (EBL) will be appreciable. This will result in a softening of the intrinsic GRB spectrum and so the soft-source analysis, optimized on the 3.5 spectral index is warranted. Though these analyses are optimized for a specific source intensity and spectral index, the detection of a source with characteristics significantly different than those assumed is not precluded.

A search for VHE emission associated with each GRB is performed over the entire duration of VERITAS observations. For well-localized bursts, a search over a shorter timescale that optimizes the VERITAS sensitivity to a power-law decay in time, specifically $\frac{dN}{dt} \propto t^{-1.5}$, is also carried out. The motivation behind the short-timescale analysis is to prevent a long exposure from washing out a gamma-ray signal that is only detectable in the early part of the observation. The choice of -1.5 as the temporal power-law index was made based on recent observations by the {\it Fermi}-LAT which has detected over a dozen GRBs with gamma-ray emission above 100 MeV. This high energy component is seen to persist well after the lower energy emission seen by the GBM has ceased though the high energy emission itself shows little spectral evolution. The spectrum is observed to follow a power-law with an index harder than the $\beta$ obtained from the Band function fit to the GBM data \cite{Ghisellini:2010p2000}. The long-lived, high-energy emission seen from the brightest LAT-detected GRBs shows a common power-law decay in time with temporal index between -1.2 and -1.7 and so the choice of -1.5 as a typical high energy temporal decay index is reasonable. This assumed temporal behavior defines an optimum duration over which VERITAS observations will be maximally sensitive. The optimum duration is strongly influenced by the delay from the start of the GRB to the beginning of VERITAS observations. For a VERITAS observation delay of 100 s, the optimum observation duration is $\sim 2 - 5$ minutes for GRBs similar to the brightest LAT-detected bursts. 

 
In one VERITAS-observed burst, GRB\,080310, the {\it Swift}-XRT observed strong X-ray flare activity persisting for thousands of seconds after the start of the burst. One very large X-ray flare was detected $\sim 475$ s after the beginning of the GRB and this flare was completely covered by the VERITAS observations, which began $342$ s after the start of the burst. A search for VHE emission coincident with this large flare, over the timeframe $475 < t-T_{0} < 750$, where $T_0$ is the start of the burst as measured by the BAT, is performed. 

\section{Results \& Discussion}

An analysis of VERITAS data associated with all well-localized GRBs shows no significant gamma-ray signal over the entire duration of VERITAS observations. The significance distributions over the field-of-view associated with both the standard and soft-source analyses are consistent with the distributions expected if no gamma-ray signal is present. The $99\%$ confidence-level upper limits on the number of VHE gamma rays, computed for each GRB using the method of \cite{2005NIMPA.551..493R}, translate to upper limits on ${\nu}$F$_{\nu}$ in the range $10^{-11} - 10^{-12}$ erg cm$^{-2}$ s$^{-1}$ at the energy threshold of each burst, which ranges from 120 GeV to 1.3 TeV depending on the elevation of observation and whether the standard or soft-source analysis was used \cite{VTSGRBsub}.  For the GRBs only detected by the GBM, no obvious source of gamma rays is found in the FOV of VERITAS observations. Analysis of these bursts is ongoing and the full results will be presented at the meeting.

 The search for gamma-ray emission on timescales optimized on the VERITAS sensitivity to a GRB with temporal characteristics similar to the bright LAT-detected bursts discussed in the previous section also yielded no significant detection. This search was only performed on 16 {\it Swift}-detected GRBs in the 2007 -- 2009 timeframe and analysis of more recent {\it Swift}-localized GRBs as well as GBM-detected bursts is ongoing. It should be noted that, though the bursts detected solely by the GBM have poor localization, they are still of interest as VERITAS targets due to the high energy range of the GBM relative to the detectors on-board {\it Swift}. 
 
No VHE gamma-ray emission associated with the large flare in GRB\,080310 was detected. After accounting for the effect of VHE gamma-ray absorption by the EBL using the model of \cite{2009MNRAS.399.1694G}, an integral upper limit of  $1.1 \times 10^{-8}$ ($3.1 \times 10^{-7}$) ph cm$^{-2}$ s$^{-1}$ above $310$ ($400$) GeV using the soft (standard) source analysis is obtained. The flare was quite intense in the XRT band, increasing by $\sim 3$ orders of magnitude relative to the underlying afterglow, but the absorption of VHE gamma-rays by the EBL for this burst, which was detected at a redshift of $z=2.4$, is also quite significant. A search for VHE emission coincident with other VERITAS-observed GRB X-ray flares is in progress.

\section{Future Prospects}

 \begin{figure*}[th]
  \centering
  \includegraphics[scale=0.9]{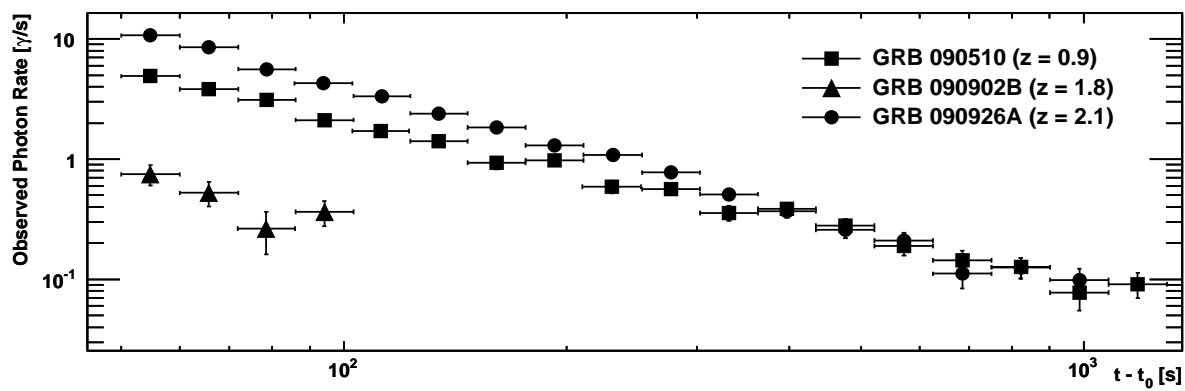}
  \caption{Predicted VERITAS lightcurves for three of the four brightest {\it Fermi}-LAT GRBs. The
    fourth, GRB\,080916C had a redshift of nearly 4.4 and VHE emission is predicted to be too
    attenuated by the EBL to be detectable by VERITAS. The EBL model of
    \cite{2009MNRAS.399.1694G} is used to estimate the attenuation of the VHE $\gamma$-rays. The
    elevation of the burst with respect to VERITAS is chosen to be 70\degree and no intrinsic
    spectral cutoff of the high energy emission is assumed. Each point signifies a detection of at
    least three standard deviations ($\sigma$) in that time bin. Figure from \cite{VTSGRBsub}.}
  \label{pred_grb_lc}
 \end{figure*}

In light of the observations by Fermi-LAT it is clear that the detection of a VHE signal from GRBs remains a difficult, but not unfeasible challenge. The duty cycle of VERITAS and other IACTs is only 10 -- 15\% and the LAT detects only about one of every fifteen GBM-detected GRBs, giving a {\it Fermi}-LAT GRB detection rate of approximately one every couple of months. The probability of a simultaneous LAT detection and VHE observation is not very high. However, as mentioned above, VERITAS did observe the LAT-detected GRB 110428A though with the delay of the distribution of the LAT burst location combined with the relatively poor localization of the GBM position it was later apparent that the GRB was not in the field that was observed with VERITAS.

Though LAT-detected GRBs are rare, their spectral and temporal characteristics lend themselves favorably to detection by ground-based IACTs like VERITAS. The high-energy GRB emission detected by the LAT is substantially delayed with respect to the prompt lower-energy burst and it is seen that, at least for the more energetic of the LAT-detected bursts, the spectral index is also harder than that found from the Band-function fit at energies below a few tens of MeV.

To make a more precise estimate of what an instrument like VERITAS may see under good observing conditions from a LAT-detected GRB, we take the characteristics of the four brightest bursts seen by the LAT: GRB 080916C \cite{AbdoSci2009}, GRB 090510 \cite{PasqualeApJL2010}, GRB 090902B \cite{AbdoApJL2009}, and GRB 090926A \cite{AckermannApJ2011} and extrapolate in both time and energy. Figure 2 shows the predicted VERITAS lightcurve from three of the four GRBs considered. In each case the high-energy spec-tral index observed by the LAT is extrapolated to the VERITAS energy band (assuming an elevation angle of 70$\degree$). The EBL attenuation of the VHE GRB spectrum is accounted for, again using the model of [14] but no intrinsic cut-off on the GRB spectrum is assumed. The fourth burst, GRB 080916C was too far away (z $>$ 4) to be detected, due to absorption of VHE gamma-rays by the EBL. Provided the intrinsic GRB spectrum extends up to VHE energies, it is apparent that some GRBs may be easily detectable by VERITAS, even up to many hundreds of seconds. Even the short burst GRB 090510, from which emission was detected by the LAT for $\sim$ 200 s, could be detected by VERITAS. It is evident, though, that follow-up observations must be made as quickly as possible as the window for observing detectable emission may be very small, barring any late-time emission e.g. that associated with X-ray flares or activity in the forward shock.

\section{Acknowledgments}
This research is supported by grants from the US Department of Energy, the US National Science Foundation, the Smithsonian Institution, by NSERC in Canada, by Science Foundation Ireland, and by STFC in the UK. We acknowledge the excellent work of the technical support staff at the FLWO and the collaborating institutions in the construction and operation of the instrument. T.~A. acknowledges support from a NASA GSRP fellowship.


\clearpage


\begin{thebibliography}{}

\bibitem{1999PhR...314..575P}
{Piran}, T. Phys. Rep. 1999, {\bf 314}, 575

\bibitem{2005ApJ...633.1018P}
{Pe'er}, A., \& {Waxman}, E. ApJ, 2005, {\bf 633}, 1018

\bibitem{2008MNRAS.384.1483F}
{Fan}, Y., {Piran}, T., {Narayan}, R., \& {Wei}, D. MNRAS, 2008, {\bf 384}, 1483

\bibitem{2009ApJ...703...60X}
{Xue}, R.~R., {et~al.} ApJ, 2009, {\bf 703}, 60

\bibitem{2007ApJ...671.1903C}
{Chincarini}, G., {et~al.} ApJ, 2007, {\bf 671}, 1903

\bibitem{FermiFlare}
{Abdo}, A.~A., {et~al.} ApJL, 2011, {\bf 734}, L27

\bibitem{2009arXiv0912.3841P}
{Perkins}, J.~S., {Maier}, G., \& {The VERITAS Collaboration}. 2009, ArXiv
  e-prints, 0912.3841
  
\bibitem{2008ICRC....3.1385C}
{Cogan}, P. 2008, in Proc. XXX Int. Cosmic Ray Conf., Vol.~3, M\'{e}rida,
  M\'{e}xico, ed. {Caballero}, R. et al., 1385--1388
  
\bibitem{Kaneko:2006p3159}
Kaneko, Y., {et~al.} ApJ Supp., 2006, {\bf 166}, 298

\bibitem{Ghisellini:2010p2000}
Ghisellini, G., {et~al.} MNRAS, 2010, {\bf 403}, 926

\bibitem{VTSGRBsub}
Acciari, V.~A., {et~al.} ApJ, 2011 (submitted)

\bibitem{Evans:2007p2698}
Evans, P.~A., {et~al.} A\&A, 2007, {\bf 469}, 379

\bibitem{Evans:2009p2640}
---. MNRAS, 2009, {\bf 397}, 1177

\bibitem{2005NIMPA.551..493R}
{Rolke}, W.~A., {et~al.} NIM A, 2005, {\bf 551}, 493

\bibitem{2009MNRAS.399.1694G}
{Gilmore}, R.~C., {et~al.} MNRAS, 2009, {\bf 399}, 1694

\bibitem{AbdoSci2009}
{Abdo}, A.~A., {et~al.} Science, 2009, {\bf 323}, 1688

\bibitem{PasqualeApJL2010}
{Pasquale}, M.~D., {et~al.} ApJL, 2009, {\bf 709}, L146

\bibitem{AbdoApJL2009}
{Abdo}, A.~A., {et~al.} ApJL, 2009, {\bf 706}, L138

\bibitem{AckermannApJ2011}
{Ackermann}, M., {et~al.} ApJ, 2011, {\bf 729}, 114



\end{thebibliography}
\end{document}